\documentclass[aps,superscriptaddress,twocolumn,10pt,a4paper,nofootinbib]{revtex4}
\usepackage[utf8]{inputenc}
\usepackage[english]{babel}

\usepackage{amssymb,amsmath,amsfonts,amsthm}
\usepackage{verbatim}

\usepackage{microtype}

\usepackage{xcolor}
\usepackage{graphicx}
\usepackage[caption=false]{subfig}
\usepackage[bookmarks,colorlinks]{hyperref}
\usepackage{url}
\usepackage{verbatim}

\definecolor{dred}{rgb}{.8,0.2,.2}
\definecolor{ddred}{rgb}{.8,0.5,.5}
\definecolor{dblue}{rgb}{.2,0.2,.8}
 
\newcommand{\bra}[1]{\langle#1|}
\newcommand{\ket}[1]{|#1\rangle}

\newcommand{\outprod}[2]{\ket{#1}\bra{#2}}

\newcommand{\half}{\mbox{$\textstyle \frac{1}{2}$}}

\newcommand{\brakets}[2]{\langle\, #1\,|\,#2\,\rangle}
\newcommand{\bracket}[3]{\left\langle #1 \left| #2 \right| #3 \right\rangle}
\newcommand{\brackets}[3]{\langle #1 | #2 | #3 \rangle}
\newcommand{\proj}[1]{\ket{#1}\bra{#1}}

\DeclareMathOperator{\Span}{span}
\DeclareMathOperator{\real}{Re}

\DeclareMathOperator{\tr}{tr}

\newcommand{\ii}{\mathrm{i}}
\newcommand{\dd}{\mathrm{d}}

\newcommand{\tave}[1]{\widehat{#1}}

\newcommand{\sym}[1]{\widetilde{#1}}

\newcommand{\E}{E}
\newcommand{\HPr}{\Lambda}

\newcommand{\SPr}{\Pi}

\newcommand{\TM}{R}

\newcommand{\T}{T}
\newcommand{\F}{F}
\renewcommand{\P}{P} 

\newcommand{\AAA}{\mathcal{A}}
\newcommand{\BBB}{\mathcal{B}}
\newcommand{\CCC}{\mathcal{C}}
\newcommand{\HHH}{\mathcal{H}}

\newcommand{\NNN}{\mathcal{N}}

\newcommand{\CC}{X}

\newcommand{\VV}{\mathcal{V}}

\newcommand{\eqr}[1]{Eq.~(\ref{#1})}
\newcommand{\fir}[1]{Fig.~\ref{#1}}

\begin{document} 

\title{Community Detection in Quantum Complex Networks}

\author{Mauro Faccin}\email{mauro.faccin@isi.it}
\affiliation{ISI Foundation, Via Alassio 11/c, 10126 Torino, Italy}
\author{Piotr Migda{\l}}
\affiliation{ICFO--Institut de Ci\`{e}ncies Fot\`{o}niques, 08860 Castelldefels (Barcelona), Spain}
\affiliation{ISI Foundation, Via Alassio 11/c, 10126 Torino, Italy}
\author{Tomi H.~Johnson}
\affiliation{Centre for Quantum Technologies, National University of Singapore, 3 Science Drive 2, 117543, Singapore}
\affiliation{Clarendon Laboratory, University of Oxford, Parks Road, Oxford OX1 3PU, United Kingdom}
\affiliation{Keble College, University of Oxford, Parks Road, Oxford OX1 3PG, United Kingdom}
\affiliation{ISI Foundation, Via Alassio 11/c, 10126 Torino, Italy}
\author{Ville Bergholm}
\affiliation{ISI Foundation, Via Alassio 11/c, 10126 Torino, Italy}
\author{Jacob D.~Biamonte}
\affiliation{ISI Foundation, Via Alassio 11/c, 10126 Torino, Italy}

\begin{abstract}
Determining community structure is a central topic in the study of complex
networks,
be it technological, social, biological or chemical, in static or interacting
systems.
In this paper, we extend the concept of community detection from classical to quantum systems
--- a crucial missing component of a theory of complex networks based on quantum mechanics. 
We demonstrate that certain quantum mechanical effects 
cannot be captured using current classical complex network tools, and
provide new methods that overcome these problems.
Our approaches are based on defining closeness measures between nodes,
and then maximizing modularity with hierarchical clustering.
Our closeness functions are based on quantum transport probability and
state fidelity, two important quantities in quantum information theory.
To illustrate the effectiveness of our approach in detecting community
structure in quantum systems, we provide several examples, including a
naturally occurring light harvesting complex, LHCII.
The prediction of our simplest algorithm, semi-classical in nature,
mostly agrees with a proposed partitioning for the LHCII found in quantum chemistry literature,
whereas our fully quantum treatment of the problem of uncover a new, consistent
and appropriately quantum community structure.
\end{abstract}

\maketitle

The identification of the community structure within a network addresses the 
problem of characterizing the
mesoscopic boundary between the microscopic scale of basic network components 
(herein called nodes) and the macroscopic scale of the whole 
network~\cite{girvan2002,porter2009communities,fortunato2010community}. 
In non-quantum networks, the detection of community structures dates back to 
Rice~\cite{rice1927identification}. 
Such analysis has revealed countless important hierarchies of
community groupings within real-world complex networks.
Salient examples can be found in social networks such as
human~\cite{zachary1977information} or animal relationships~\cite{lusseau2004}, 
biological~\cite{jonsson2006cluster,pimm1979structure,krause2003compartments,
  guimera2005functional},
biochemical~\cite{holme2003subnetwork} and
technological~\cite{flake2002self,gauvain2013communities} networks, 
 as well as numerous others (see Ref.~\cite{girvan2002}).
In quantum networks, as researchers explore networks of an increasingly 
non-trivial geometry and large 
size~\cite{allegra2012,plenio2008dephasing,renger2006, baezbook}, their analysis and understanding will involve identifying non-trivial community structures. 
In this article, we devise methods to perform this task, providing 
an important missing component in the recent drive to unite quantum physics and
complex network science \cite{faccin2013degree, zimboras2013quantum}.

For quantum systems, beyond being merely a tool for analysis following
simulations, community partitioning is closely related to performing the
simulations themselves. Simulation is generally a difficult task
\cite{johnson2014quantum}, e.g.\ simulating 
exciton transport in dissipative quantum biological networks 
\cite{ringsmuth2012,fleming10,IF12,CF09,caruso09,MRLA08}. 
The amount of resources required to exactly simulate such processes scales 
exponentially with the number of nodes. 
To overcome this, one must in general seek to describe only limited
correlations between certain parts of the network \cite{banchi2013foster}. Mean-field
\cite{rokhsar1991gutzwiller,krauth1992gutzwiller,sheshadri1993superfluid,luhmann2013bosonic} and
tensor-network methods \cite{verstraete2008matrix,cirac2009renormalization} assume correlations between bi-partitions of the
system along some node-structure to be zero or limited by an area law.
Hartree-Fock methods assume limited correlations between particles
\cite{wang2011numerically,cao2013multilayer}. Thus planning a simulation involves identifying a
partitioning of a system for which it is appropriate to limit
inter-community correlations, i.e.\ is a type of community detection.

We apply our detection methods to artificial networks and the real-world light harvesting
complex~II (LHCII) network.
In past works, researchers have divided the LHCII \emph{by hand} in order to gain more insight into the
system dynamics~\cite{pan2013architercture,novoderezhkin2005lhcii,fleming2009lhcii}.
Meanwhile, our methods optimize the task of identifying communities within
a quantum network \emph{ab initio} and, as we will show, the resulting
communities consistently point towards a structure that is different to those
previously identified for the LHCII\@. For larger networks, as with our artifical examples, automatic methods would appear to be the only
feasible option.

Our specific approach is to generate a hierarchical community structure 
\cite{carlsson2010characterization} by defining both inter-node and inter-community closeness.
The optimum level in the hierarchy is determined by a modularity-based measure,
which quantifies how good a choice of communities is for the quantum network
on average relative to an appropriately randomized version of the network.
Although modularity-based methods are known to struggle with large sparse
networks~\cite{guimera2004modularity,good2010performance},
this work focuses on quantum systems whose size remains much smaller than
the usual targets of classical community detection algorithms.

While the backbone of our quantum community method is shared with classical 
methods, the physical properties used to characterize a good community in a quantum must necessarily be very different to the properties used for a classical system.
Here we show how two quantum properties are used to 
obtain closeness and modularity functions: the first is the coherent transport 
between communities and the second is the 
change in the states of individual communities during a coherent
evolution.

\begin{figure}[b!]
  \centering
  \begin{center}
      \includegraphics[width=\columnwidth]{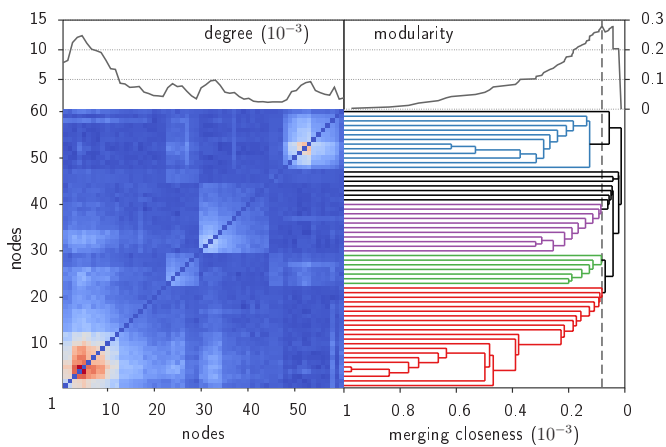}
  \end{center}
\caption{Hierarchical community structure arising from a quantum evolution. 
On the left are the closenesses $c(i,j)$ between $n=60$ nodes.
On the right is the dendrogram showing the resulting hierarchical community 
structure.
The dashed line shows the optimum level within this hierarchy, according to the 
modularity.
The particular example shown here is the one corresponding 
to Fig.~\ref{fig:artificial}d.
}
\label{fig:dendro}
\end{figure}

In Section~\ref{sec:comdet} we will begin by recalling several common notions from classical community
detection that we rely on in this work.  This sets the stage for the
development of a quantum treatment of community detection in Section~\ref{sec:quantumcom}. 
We then turn to several examples in Section~\ref{sec:performance} including
the LHCII complex mentioned previously, before concluding in Section~\ref{sec:discussion}.
\section{Community detection}\label{sec:comdet}

Community detection is the partitioning of a set
of nodes~$\NNN$
into non-overlapping~\footnote{We do not consider 
  generalizations to overlapping communities here.}
and non-empty subsets 
$\AAA,~\BBB,~\CCC,\ldots~\subseteq~\NNN$,
called communities, that together cover $\NNN$.

There is usually no agreed upon optimal partitioning of nodes into communities. 
Instead there is an array of approaches that differ in both the definition of 
optimality and the method used to achieve, exactly or approximately, this 
optimality (see Ref.~\cite{fortunato2010community} for a recent review). In classical 
networks optimality is, for example, defined 
statistically~\cite{lancichinetti2011oslom}, e.g.\ in terms of 
connectivity~\cite{girvan2002} or 
communicability~\cite{estrada2011community,estrada2009community}, or increasingly, and sometimes 
relatedly~\cite{meila2001random}, in terms of stochastic random 
walks~\cite{Delvenne2010,rosvall2011infomap,pons2005computing}. Our particular focus is on the 
latter, since the concept of transport (e.g.~a quantum walk) is central to
nearly all studies conducted in quantum physics.  As for achieving optimality,
methods include direct maximization 
via simulated annealing~\cite{guimera2004modularity,guimera2005functional} or, 
usually faster, iterative division or agglomeration of 
communities~\cite{hastie2001elements}. 
We focus on the latter since it provides a simple and effective 
way of revealing a full hierarchical structure of the network, requiring only 
the definition of the closeness of a pair of communities.

Formally, hierarchical community structure detection methods are
based on a (symmetric) closeness function
$c(\AAA,\BBB) = c(\BBB,\AAA)$
of two communities $\AAA \neq \BBB$. 
In the agglomerative approach,
at the lowest level of the hierarchy, the nodes are
each assigned their own communities. An iterative procedure then
follows, in each step of which the closest pair of communities
(maximum closeness $c$) are merged. This procedure ends at the highest level, where
all nodes are in the same community.
To avoid instabilities in this agglomerative procedure, the closeness
function is required to be non-increasing under the merging of two communities,
$c(\AAA \cup \BBB, \CCC) \le \max(c(\AAA,\CCC), c(\BBB,\CCC))$,
which allows the representation of the community
structure as a linear hierarchy indexed by the merging closeness.
The resulting structure is often represented as a dendrogram (as shown in 
\fir{fig:dendro})%
~\footnote{%
In general it may happen that more than one pair of communities are at the 
maximum closeness. In this case the decision on which pair merges first can 
influence the structure of the dendrogram, 
see~\cite{jain1988algorithms,carlsson2010characterization}.
In~\cite{carlsson2010characterization} a permutation invariant formulation of 
the agglomerative algorithm is given, where more than two clusters can be merged 
at once. In our work we use this formulation unless stated otherwise.
}.

This leaves open the question of which level of the hierarchy yields
the optimal community partitioning. If a
partitioning is desired for simulation, for example, then there may be
a desired maximum size or minimum number of communities. However, without such
constraints, one can still ask what is the best choice of communities
within those given by the hierarchical structure.

A type of measure that is often used to quantify the quality of a community partitioning
choice for this purpose is
modularity~\cite{newman2004pre,Newman2004,newman2004finding}, denoted $Q$.
It was originally
introduced in the classical network setting, in which a network is
specified by a (symmetric) adjacency matrix of (non-negative) elements
$A_{ij} = A_{ji} \ge 0$ ($A_{ii} = 0$), each off-diagonal element giving the weight of
connections between nodes $i$ and $j\neq i$~\footnote{As will become apparent, we need 
  only consider undirected networks without self-loops.}.
The modularity attempts to
measure the fraction of weights connecting elements in the same
community, relative to what might be expected. Specifically, one takes the fraction of intra-community
weights and subtracts the average fraction obtained when the
start and end points of the connections are reassigned randomly,
subject to the constraint that the total connectivity $k_i = \sum_j
 A_{ij}$ of each node is fixed. The modularity is then given
by
\begin{align}
\label{eq:modularity}
  Q = \frac{1}{2m} \tr \left \{ C^{\mathrm{T}} B C \right \} ,
\end{align}
where $m = \half \sum_i k_i$ is the total weight of connections, $B$
is the modularity matrix with elements $B_{ij} = A_{ij}-k_i k_j / 2 m$, and
$C$ is the community matrix, with elements $C_{i \AAA}$ equal to unity if $i \in
\AAA$, otherwise zero. 
The modularity then takes values strictly less than one, possibly negative, and exactly zero in the case that the nodes form a single community.

As we will see, there is no natural adjacency matrix associated with
the quantum network and so for the purposes of modularity we use
$A_{ij} = c (i , j)$ for $i \neq j$. The modularity $Q$ thus measures
the fraction of the closeness that is intra-community, relative to
what would occur if the inter-node closeness $c (i , j)$ were randomly
mixed while fixing the total closeness $k_i = \sum_{j\neq i} c (i ,
j)$ of each node to all others. Thus both the community structure and
optimum partitioning depend solely on the choice of the closeness
function.

  Modularity-based methods such as above are intuitive, fast and on the most part effective,
  yet we must note that for classical systems it has been shown that modularity-based methods suffer from a number
  of flaws that influence the overall efficacy of those
  approaches.
  In Refs.~\cite{guimera2004modularity,good2010performance} modularity-based
  methods show a poor performance in large, sparse real-world and model 
  networks. This is due mainly to the resolution limit
  problem~\cite{fortunato2007resolution}, where small communities
  can be overlooked, and modularity landscape degeneracy, which
  strongly influence accuracy in large networks.
  Another modularity-related problem is the so-called 
  detectability/undetectability 
  threshold~\cite{nadakuditi2012graph,radicchi2013detectability,radicchi2014paradox}
  where an approximate bi-partition of the system becomes undetectable in some cases, in
  particular in presence of degree homogeneity.
However, in the present work we focus on quantum networks whose size 
typically remains small compared to classical targets of community
detection algorithms,
and for which the derived adjacency matrices are not sparse.
These characteristics help to limit the known flaws of our
modularity-based approach, making it adequate for our purposes.

Finally, once a community partitioning is obtained it is often desired
to compare it against another. Here we use the common 
normalized mutual information
(NMI)~\cite{ana2003normalized,strehl2003cluster,danon2005}
as a measure of the mutual dependence of two community partitionings.
Each partitioning
$\CC = \{\AAA, \BBB,\dots \}$
is represented by a probability distribution
${P_\CC = \{|\AAA|/|\NNN|\}_{\AAA \in \CC}}$, where
$|\AAA| = \sum_i C_{i \AAA}$ is the
number of nodes in community~$\AAA$. The similarity of two community
partitionings $\CC$ and $\CC'$ depends on the joint
distribution
$P_{\CC \CC'} = \{ |\AAA \cap \AAA'| / |\NNN| \}_{\AAA \in \CC, \AAA' \in \CC'}$,
where
$|\AAA \cap \AAA'| = \sum_i C_{i \AAA} C_{i \AAA'}$
is the number of nodes that belong to both communities $\AAA$ and $\AAA'$. Specifically, NMI is defined as
\begin{equation}\label{eq:nmi}
\operatorname{NMI}(\CC,\CC') = 
\frac{2\,I(\CC,\CC')}{H(\CC)+H(\CC')}.
\end{equation}
Here $H(\CC)$ is the Shannon entropy of $P_\CC$, and the
mutual information $I(\CC,\CC')=
H(\CC)+H(\CC')-H(\CC,\CC')$ depends on the
entropy $H(\CC,\CC')$ of the joint distribution
$P_{\CC \CC'}$.
The mutual information is the average of the amount of information about
the community of a node in $\CC$ obtained by learning its
community in $\CC'$.
The normalization
ensures that the NMI has a minimum value of zero and takes its maximum value of unity for two identical
community partitionings. 
The symmetry of the definition of NMI follows from that of mutual
information and Eq.~\eqref{eq:nmi}.
  
\section{Quantum community detection}\label{sec:quantumcom}
The task of community detection has a particular interpretation in a quantum
setting. The state of a quantum system is described in terms of a
Hilbert space $\HHH$, spanned by a complete orthonormal set of basis
states $\{ \ket{i} \}_{i \in \NNN}$. Each basis state~$\ket{i}$ can be associated
with a node~$i$ in a network and often, as in the case of single
exciton transport, there is a clear choice of basis states that makes
this abstraction to a spatially distributed network natural.

The partitioning of nodes into communities then corresponds to the
partitioning of the Hilbert space $\HHH = \bigoplus_{\AAA \in \CC} \VV_\AAA $
into mutually orthogonal subspaces $\VV_\AAA = \Span_{i \in \AAA} \{\ket{i} \}$.
As with classical networks, one can then imagine an assortment of
optimality objectives for community detection, for example, to identify a
partitioning into subspaces in which inter-subspace transport is
small, or in which the state of the system remains relatively unchanged within each subspace.
In the next two subsections we introduce two classes of community closeness
measures that correspond to these objectives.
Technical details can be found in the Supplemental Material.

In what follows, we focus our analysis one an isolated quantum
system governed by Hamiltonian~$H$, which enables us to derive
convenient closed-form expressions for the closeness measures.
We may expand~$H$ in the node basis $\{ \ket{i} \}_{i \in \NNN}$:
\begin{equation}
    H = \sum_{ij}  H_{ij} \ket{i}\bra{j} .
\end{equation}
A diagonal element $H_{ii}$ is a real value denoting the energy of state $\ket{i}$, whilst an off-diagonal element $H_{ij}$, $i \neq j$, is a complex
weight denoting the change in the amplitude of the wave function during a transition from state $\ket{j}$ to $\ket{i}$. The matrix formed by these elements 
can be thought of as a $|\NNN| \times |\NNN|$ complex, hermitian adjacency matrix.
In quantum mechanics, complex elements in the
Hamiltonian lead to a range of phenomena not captured by real matrices, such as time-reversal symmetry 
breaking~\cite{zimboras2013quantum,lu2014chiral}.
In the case where each state~$\ket{i}$ corresponds to a particle being localised at a spatially distinct node~$i$,
the Hamiltonian describes a spinless single-particle walk with an energy landcape given by the diagonal elements, and
transition amplitudes by the off-diagonal elements.
Any quantum evolution can be viewed in this picture, making the single particle spiness walk scenario rather general.  

A community partitioning based on a Hamiltonian $H$ could be 
used, among other things, to guide the simulation or analysis of a more complete
model in the presence of an environment, where this more complete
model may be much more difficult to describe.
Additionally, our method could be generalized to use closeness measures
based on open-system dynamics obtained numerically.

\subsection{Inter-community transport}
\label{sec:mixing}
Several approaches to detecting communities in classical networks are
based on the flow of probability through the network during a
classical random walk~\cite{meila2001random,eriksen2003modularity,pons2005computing,weinan2008optimal,Delvenne2010,rosvall2011infomap}.
In particular, many of these methods seek communities for which the
inter-community probability flow or transport is small. 
A natural approach to quantum community detection is thus to
consider the flow of probability during a continuous-time quantum
walk, and to investigate the \emph{change} in the probability of observing
the walker within each community:
\begin{align}
\T_{\CC}(t) &= \sum_{\AAA \in \CC} \T_\AAA (t)
= \sum_{\AAA \in \CC} \frac{1}{2}\left| p_\AAA \left \{ \rho (t) \right \} - p_\AAA \left \{ \rho (0) \right \} \right|,
\end{align}
where
$
\rho (t) = \mathrm{e}^{-\ii H t} \rho (0) \mathrm{e}^{\ii H t}
$
is the state of the walker, at time~$t$, during the walk generated by $H$, and
\begin{align}
p_\AAA \left \{ \rho \right \} = \tr \left \{ \SPr_\AAA \rho \right \} ,
\end{align}
is the probability of a walker in state~$\rho$ being found in
community~$\AAA$ upon a von Neumann-type measurement~\footnote{Equivalently, $p_\AAA \left \{ \rho \right \}$ is
the norm of the projection (performed by projector $\SPr_\AAA$) of the
state $\rho$ onto the community subspace $\VV_\AAA$.}.
$\SPr_\AAA=\sum_{i\in\AAA} \ket{i}\bra{i}$ denotes the projector to the
$\AAA$ subspace.

The initial state~$\rho (0)$ can be chosen freely.
The change in inter-community transport is clearest when the process begins either entirely inside or entirely outside each community. Because of this, we choose the walker to be initially localized
at a single node $\rho (0) = \proj{i}$ and then, for symmetry, sum
$\T_\CC (t)$ over all $i \in \NNN$. This results in the particularly
simple expression
\begin{align}
\T_\AAA (t) = \sum_{i \in \AAA, j \notin \AAA} \frac{\TM_{ij}(t)+\TM_{ji}(t)}{2}
= \sum_{i \in \AAA, j \notin \AAA} \sym{\TM}_{ij}(t),
\end{align}
where $\TM(t)$ is the doubly stochastic transfer matrix whose elements
$\TM_{ij}(t) = |\brackets{i}{\mathrm{e}^{-\ii H t}}{j}|^2$ give
the probability of transport from node~$j$ to node~$i$,
and $\sym{\TM}(t)$ its symmetrization.
This is reminiscent of classical community detection methods,
e.g.~\cite{pons2005computing}, using closeness measures based on the
transfer matrix of a classical random walk.

We can thus build a community structure that seeks to reduce $\T_\CC (t)$
at each hierarchical level by using the closeness function
\begin{align}
\label{eq:closeness_transport}
\notag
c^\T_t(\AAA , \BBB) &= \frac{\T_\AAA (t) + \T_\BBB (t) -
  \T_{\AAA\cup\BBB}(t)}{|\AAA||\BBB|}\\
&= 
\frac{2}{|\AAA||\BBB|} \sum_{i \in \AAA, j \in \BBB} \sym{\TM}_{ij}(t) ,
\end{align}
where the numerator is the decrement in $\T_\CC (t)$ caused by merging
communities $\AAA$ and $\BBB$.
The normalizing factor in Eq.~\eqref{eq:closeness_transport} avoids the
effects due to the uninteresting scaling of the numerator with the community
size.

Since a quantum walk does not converge to a stationary state, a
time-average of the closeness defined in Eq.~\eqref{eq:closeness_transport}
is needed to obtain a quantity that eventually converges with
increasing time. 
Given the linearity of the formulation, this corresponds to replacing
the transport probability~$\TM_{ij}(t)$
in \eqr{eq:closeness_transport}
with its time-average
\begin{align}
\label{eq:avtransfer}
\tave{\TM}_{ij}(t) = \frac{1}{t} \int_0^t \TM_{ij}(t') \:\dd t'.
\end{align}

It follows that, as with similar classical community detection methods~\cite{Delvenne2010},
our method is in fact a class of approaches, 
each corresponding to a different time $t$. 
The appropriate value of $t$ will depend on the specific
application, for example, a natural
time-scale might be the decoherence time. 
Not wishing to lose generality and focus on a particular system, 
we focus here on the short and long time limits. 

In the short time limit $t \to 0$, relevant if $t H_{ij} \ll 1$ for
$i \neq j$, the averaged transfer matrix $\tave{T}_{ij} (t)$ is simply
proportional to $|H_{ij}|^2$.
Note that in the short
time limit there is no interference between
different paths from $\ket{i}$ to $\ket{j}$, and therefore for
short times $c^\T_t (i , j)$ does not depend on the on-site energies $H_{ii}$ or the phases of the
hopping elements $H_{i j}$.
This is because, to leading order in time, interference
does not play a role in the transport out of a single node. 
For this reason we can refer to this approach as ``semi-classical''.

In the long time limit $t \to \infty$, relevant if $t$ is much larger
than the inverse of the smallest gap between distinct
eigenvalues of $H$, the probabilities are elements of the
mixing matrix~\cite{godsil2011},
\begin{align}
\lim_{t\to \infty} \tave{\TM}_{ij}(t) = \sum_k | \bracket{i}{\HPr_k}{j}|^2 ,
\label{eq:mixing-matrix}
\end{align}
where $\HPr_k$ is the projector onto the $k$-th eigenspace of $H$. This
thus provides a simple spectral method for building the community
structure.

Note that, unlike in a classical infinitesimal stochastic walk where
each $\tave{\TM}_{ij} (t)$
eventually becomes proportional to the connectivity $k_j$ of the final
node $j$, the long time limit in the quantum setting is non-trivial and,
as we will see, $\tave{\TM}_{ij}(t)$ retains a strong impression of the
community structure for large~$t$%
~\footnote{Note that, apart from small or large
times $t$, there is no guarantee of symmetry $\TM_{ij}(t) = \TM_{ji}(t)$ in the
transfer matrix for a given
Hamiltonian. See~\cite{zimboras2013quantum}. Hamiltonians featuring
this symmetry, e.g., those with real $H_{ij}$, are called
time-symmetric.}.

\subsection{Intra-community fidelity}
\label{sec:coher}
Classical walks, and the community detection methods based on them, are fully 
described by the evolution of the probabilities of the walker occupying each 
node. The previous quantum community detection approach is based on the 
evolution of the same probabilities but for a quantum walker.
However, quantum walks are richer than this, they are not fully
described by the evolution of the node-occupation probabilities. We
therefore introduce another community detection method that captures
the full quantum dynamics within each community subspace.

Instead of reducing merely the change in probability within the
community subspaces, we reduce the change in the projection of the
quantum state in the community subspaces. This change is measured
using (squared) fidelity, a common measure of distance between two
quantum states.
For a walk beginning in state $\rho (0)$ we therefore focus on the quantity
\begin{align}
  \F_\CC (t) &= \sum_{\AAA \in \CC} \F_\AAA (t)
= \sum_{\AAA \in \CC}  F^2 \left \{ \SPr_\AAA \rho(t) \SPr_\AAA, \SPr_\AAA \rho(0) \SPr_\AAA \right \},
\end{align}
where $\SPr_\AAA \rho \SPr_\AAA$ is the projection of the state $\rho$ onto the subspace $\VV_\AAA$ and
\begin{align}
  F \left\{\rho ,\sigma \right\} = \tr \left\{ \sqrt{\sqrt{\rho} \sigma \sqrt{\rho}} \right\} \in  [0, \sqrt{\tr \{\rho\} \tr \{\sigma\}}]
\end{align}
is the fidelity, which is symmetric between $\rho$ and $\sigma$.

We build a community structure that seeks to maximize 
the increase in $\F_\CC (t)$ at 
each hierarchical level by using the closeness measure
\begin{align}
\label{eq:fidelitydist}
c^\F_t (\AAA , \BBB) = \frac{\F_{\AAA \cup \BBB}(t) -\F_\AAA(t) -\F_\BBB(t)
}{|\AAA| |\BBB|} \in [-1 ,1],
\end{align}
i.e.\ the change in $\F_\CC (t)$ caused by merging communities $\AAA$ and~$\BBB$.
Our choice for the denominator prevents uninteresting size scaling, 
as in Eq.~\eqref{eq:closeness_transport}.

The initial state~$\rho(0)$ can be chosen freely. Here we choose the
pure uniform superposition state $\rho(0)=\ket{\psi_0}\bra{\psi_0}$ satisfying
$\brakets{i}{\psi_0} = 1/\sqrt{n}$ for all~$i$.
This state was used to 
investigate the effects of the connectivity on the dynamics of a quantum walker
in~Ref.~\cite{faccin2013degree}. 

As for our other community detection approach, we consider the time-average of 
Eq.~\eqref{eq:fidelitydist}, which yields
\begin{align}
c_t^\F (\AAA,\BBB) =
\frac{2}{|\AAA||\BBB|} \sum_{i\in\AAA, j\in\BBB} 
\real(\tave{\rho}_{ij}(t)\rho_{ji}(0)),
\end{align}
where $\tave{\rho}_{ij}(t) = \frac 1t \int_0^t \dd t' \rho_{ij}(t')$.
In the long time limit, the time-average of the density matrix takes a particularly simple 
expression:
\begin{align}
  \lim_{t \to \infty} \tave{\rho}_{ij}(t) = \sum_k \HPr_k \rho_{ij}(0) \HPr_k,
\end{align}
where $\HPr_k$ is as in the previous Sec.~\ref{sec:mixing}.

The definition of community closeness given in
Eq.~\eqref{eq:fidelitydist} can exhibit negative values. 
In this case the usual definition of modularity fails~\cite{traag2009}
and one must extended it.
In this work we use the definition of modularity proposed
in Ref.~\cite{traag2009}, which coincides with Eq.~(\ref{eq:modularity}) in
the case of non-negative closeness. 
The extended definition treats negative and positive 
links separately, and tries to minimize intra-community negative
links while maximizing intra-community positive links.

\section{Performance analysis}\label{sec:performance}

To analyze the performance of our quantum community detection methods
we apply them to three different networks.
The first one (Sec.~\ref{sec:quantumnet}) is a simple quantum network, 
which we use to highlight how some intuitive notions in 
classical community detection do not necessarily transfer over 
to quantum systems.
The second example (Sec.~\ref{sec:artificial}) is an artificial quantum
network designed to exhibit a clear classical community structure, 
which we show is different from the quantum community structure obtained and 
fails to capture significant changes in this structure induced by quantum
mechanical phases on the hopping elements of the Hamiltonian.
The final network (Sec.~\ref{sec:lhcii}) is a real world quantum
biological network, describing the LHCII light harvesting complex, for which we find a
consistent quantum community structure differing from the
community structure cited in the literature.
These findings confirm that a quantum mechanical treatment of community
detection is necessary as classical and semi-classical methods
cannot be reproduce the structures that appropriately capture quantum effects.

Below we will compare quantum community structures
against more classical community structures, such the one given
by the semi-classical method based on the short time transport and,
in the case of the example of Sec.~\ref{sec:artificial},
the classical network from which the quantum network is constructed.
Additionally we use a traditional classical
community detection algorithm, OSLOM~\cite{lancichinetti2011oslom}, an algorithm based on the maximization
of the statistical significance of the proposed partitioning,
whose input adjacency matrix~$A$ must be real.
For this purpose we use the absolute
values of the Hamiltonian elements in the site basis: $A_{ij} = |H_{ij}|$. 

\subsection{Simple quantum network}
\label{sec:quantumnet} 

\begin{figure}
  \includegraphics[width=\columnwidth]{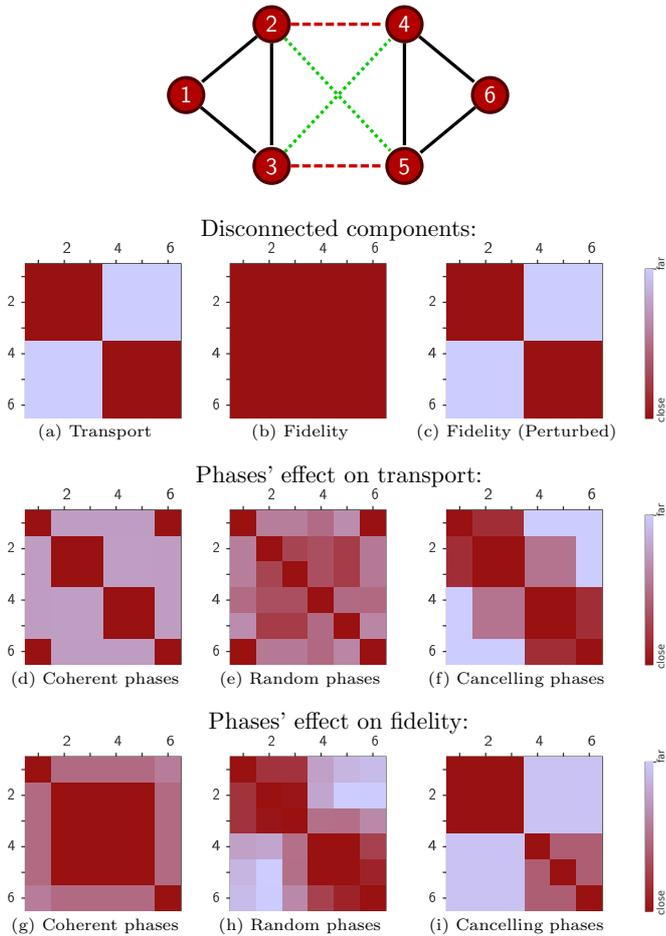}
  \caption{
  Simple quantum network --- a graph with six nodes.
  Each solid line represents transition amplitude $H_{ij}=1$.
  For dashed and dotted lines the transition amplitude can be either zero (a, b 
  and c) or the absolute value is the same $|H_{ij}|=1$ but phase is
  (d and g) coherent (all ones),
  (e and h) random $\exp(i \varphi_k)$ for each link,
  (f and i) canceling (ones for dashed red and minus one for dotted green).
  Plots show the node closeness for both methods based on transport and fidelity 
  (only the long-time-averages are considered, in plots (g), (h) and (i) we
  used a perturbed Hamiltonian to solve the eigenvalues degeneracy, this
  explains the non-symmetric closeness in (i)).
  }
  \label{fig:phase-sensitive-graph}
\end{figure}

Here we use a simple six-site network model to study ways in which quantum 
effects lead to non-intuitive results, and how methods based on different quantum 
properties can, accordingly, lead to very different choices of communities.

We begin with two disconnected cliques of three nodes each, 
where all Hamiltonian matrix elements within the groups are identical and real.
\fir{fig:phase-sensitive-graph} illustrates this highly 
symmetric topology.
The community detection method based on quantum transport identifies the two 
fully-connected groups as two separate communities 
(\fir{fig:phase-sensitive-graph}a), as is expected. 
Contrastingly, the methods based on fidelity predict
counter-intuitively only a single community; two disconnected nodes can retain 
coherence and, by this measure, be considered part of the same community 
(\fir{fig:phase-sensitive-graph}b). 

This symmetry captured by the fidelity-based community structure
breaks down if we introduce random perturbations into the Hamiltonian.
Specifically, the fidelity-based closeness~$c_t^F$
is sensitive to perturbations of the order~$t^{-1}$, above
which the community structure is divided into the two groups of three
(\fir{fig:phase-sensitive-graph}c)
expected from transport considerations. Thus we may tune the resolution of
this community structure method to asymmetric perturbations by varying~$t$.

Due to quantum interference we expect that the
Hamiltonian phases should significantly affect the quantum community partitioning.
The same toy model can be used to demonstrate this effect.
For example, consider adding four elements to the Hamiltonian corresponding to
hopping from nodes 2 and 3 to 4 and 5 (see diagram in
\fir{fig:phase-sensitive-graph}). 
If these hopping elements are all identical to the others, it
is the two nodes, 1 and 6, that are not directly connected for which the
inter-node transport is largest (and thus their inter-node closeness is the
largest). However, when the phases of the four additional elements are
randomized, this transport is decreased.
Moreover, when the phases are canceling, the
transport between nodes 1 and 6 is reduced to zero, and the closeness between
them is minimized
(see
Figs.~\ref{fig:phase-sensitive-graph}d--\ref{fig:phase-sensitive-graph}f).  

The fidelity method has an equally strong dependence on the phases (see 
Figs.~\ref{fig:phase-sensitive-graph}g--\ref{fig:phase-sensitive-graph}i), with
variations in the phases breaking up the network from a large central community
(with nodes 1 and 6 alone)
into the two previously identified communities.

\subsection{Artificial quantum network}
\label{sec:artificial}

\begin{figure*}[t!]
  \centering
  \includegraphics[width=\textwidth]{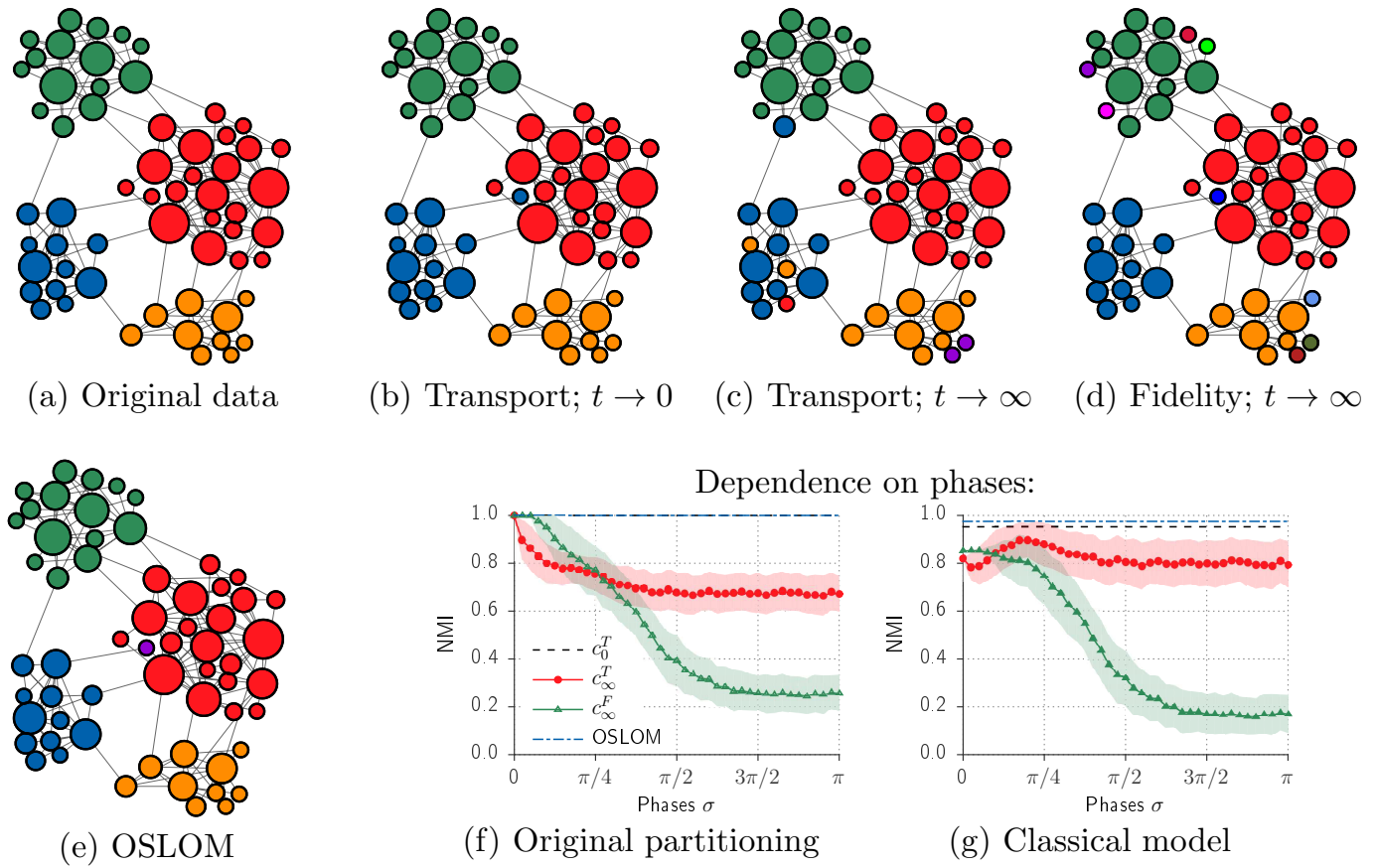}
\caption{Artificial community structure.
  (a) Classical community structure used in creating the network.
  (b--e) Community partitionings found using the
  three quantum methods and OSLOM.
  (f,g) 
  Behavior of the approaches as the phases of the Hamiltonian elements are randomly
  sampled from a Gaussian distribution of width~$\sigma$. The mean
  NMI, compared with zero phase partitioning (f) and the classical
  model data (g),
  over 200 samplings of the phase distribution is plotted. The
  standard deviation is indicated by the shading.
  Both OSLOM and $c^{\T}_0$ are insensitive to phases and thus do not
  respond to the changes in the Hamiltonian.
}
\label{fig:artificial}
\end{figure*}

The Hamiltonian of our second quantum network is constructed from the
adjacency matrix $A$ of a classical unweighted, undirected network
exhibiting a clear classical partitioning,
using the relation $H_{ij} = A_{ij}$.
We construct~$A$ using the algorithm proposed by Lancichinetti {\em et
al.}~in Ref.~\cite{lancichinetti2008}, which provides a method
to construct a network with heterogeneous distribution both for the node
degree and for the communities dimension and a controllable inter-community
connection.  We start with a rather small network of 60 nodes with average
intra-community connectivity $\langle k \rangle=6$, and only 5\% of the
edges are rewired to join communities.
The network is depicted in Fig.~\ref{fig:artificial}a.
To confirm the expected, the known classical community structure is indeed
obtained by the semi-classical short-time-transport algorithm%
~\footnote{In the case of short-time transport, a small
  perturbation was also added to the closeness
  function in order to break the symmetries of the system.}
and the OSLOM
algorithm (see Figs.~\ref{fig:artificial}b--\ref{fig:artificial}e),
achieving $\text{NMI}=0.953$ and $\text{NMI}=0.975$ with the known
structure, respectively.

The quantum methods based on the long-time average of both transport and fidelity
reproduce the main features of the original community structure
while unveiling new characteristics.  The transport-based long-time average
method ($\text{NMI}=0.82$ relative to the classical partitioning)
exhibits disconnected communities, i.e.\ the
corresponding subgraph is disconnected. This behavior can be explained
by interference-enhanced quantum walker dynamics, as exhibited by the toy
model in the previous subsection.
The
long-time average fidelity method ($\text{NMI}=0.85$) returns the four main
classical communities plus a number of single-node communities.
Both methods demonstrate that the quantum and classical community
structures are unsurprisingly different, with the quantum community
structure clearly dependent on the quantum property being
optimized, more so than the different classical partitionings.

\subsubsection*{Adjusted phases}

As shown in Sec.~\ref{sec:quantumnet},
due to interference the dynamics of the quantum system can change
drastically if the phases of the Hamiltonian elements are non-zero. This is
known as a chiral quantum walk~\cite{zimboras2013quantum}. Such walks exhibit, for example, time-reversal
symmetry breaking of transport between sites~\cite{zimboras2013quantum} and
it has been proposed that nature might actually make use of phase
controlled interference in transport processes~\cite{harel2012quantum}.
OSLOM, our semi-classical short-time transport algorithm and other
classical community partitioning methods are insensitive to changes in the
hopping phases. Thus, by establishing that the quantum community structure
is sensitive to such changes in phase, as expected from above, we show that
classical methods are inadequate for finding quantum community structure.

To analyze this effect we take the previous network
and adjust the phases of the Hamiltonian terms while preserving their
absolute values. Specifically, the phases are sampled randomly from a
normal distribution with mean zero and standard deviation $\sigma$.
We find that, typically, as the standard deviation $\sigma$ increases, 
when comparing quantum communities and the corresponding communities
without phases the NMI between them decreases, as shown in
Fig.~\ref{fig:artificial}f.
A similar deviation reflects on the comparison with the classical
communities used to construct the system, shown in
Fig.~\ref{fig:artificial}g.
This sensitivity of the quantum community structures to phases, as revealed
by the NMI, confirms the expected inadequacy of classical methods.
The partitioning based on long-time average fidelity seems to be the most sensitive
to phases.

\subsection{Light-harvesting complex}
\label{sec:lhcii} 

\begin{figure*}[t!]
\includegraphics[width=\textwidth]{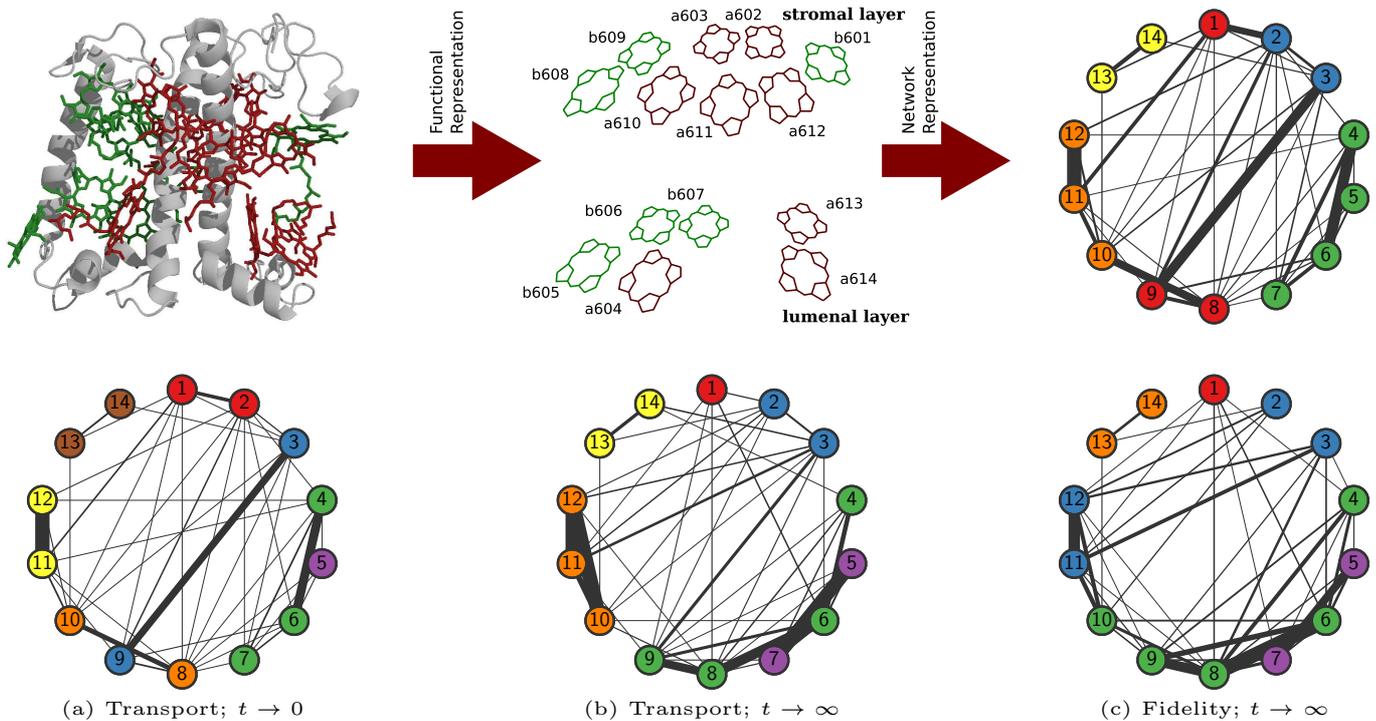}
\caption{Light harvesting complex II (LHCII).
  (top left) Monomeric subunit of the LHCII complex with pigments Chl-a (red) and 
  Chl-b (green) packed in the protein matrix (gray).
  (top center) Schematic representation of Chl-a and Chl-b in the monomeric 
  subunit, here the labeling follows the usual nomenclature (b601, a602\dots).
  (top right) Network representation of the pigments in circular layout, colors represent 
  the typical partitioning of the pigments into communities. The widths of the links 
  represent the strength of the couplings~$|H_{ij}|$ between nodes.
  Here the labels maintain only the ordering (b601$\to$1, a602$\to$2,\dots).
  (a,b,c) Quantum communities as found by the different quantum community detection methods.
  Link width denotes the pairwise closeness of the nodes.
}
\label{fig:lhcii}
\end{figure*}

An increasing number of biological networks of non-trivial  topology are
being described using quantum mechanics.  For example, light harvesting
complexes have drawn significant attention in the quantum information
community.

One of these is the LHCII, a two-layer 14-chromophore complex embedded into
a protein matrix (see Fig.~\ref{fig:lhcii} for a sketch) that
collects light energy and directs it toward the reaction center where it is
transformed into chemical energy.  The system can be described as a network
of 14 sites connected with a non-trivial topology.
The single-exciton subspace is spanned by 14
basis states, each corresponding to a node in the network, and the
Hamiltonian in this basis was found in Ref.~\cite{fleming2009lhcii}.

In a widely adopted chromophore community
structure~\cite{novoderezhkin2005lhcii}, the sites are partitioned
\emph{by hand} into communities according to their physical closeness
(e.g. there are no communities spanning the two layers of the complex), and
the strength of Hamiltonian couplings
(see the top right of Fig.~\ref{fig:lhcii}). Here, we
apply our \emph{ab initio} automated quantum community detection algorithms
to the same Hamiltonian.

All of our approaches predict a modified partitioning to that commonly used
in the literature. The method based on short-time transport returns
communities that do not connect the two layers. 
This semi-classical approach relies only on
the coupling strength of the system, without considering interference
effects, and provides the closest partitioning to the one provided by the literature
(also relying only on the coupling strengths).
Meanwhile, the methods
based on the long-time transport and fidelity return very similar community
partitionings, in which node 6 on one layer and node 9 on the other are in
the same community.
These two long-time community partitionings are
identical, except one of the communities predicted by the fidelity based
method is split when using the transport based method. It is therefore a
difference in modularity only. 

The classical OSLOM algorithm fails spectacularly: it gives only one significant community
involving nodes 11 and 12 which exhibit the highest coupling strength. If
assigning a community to each node is forced, a unique community with all
nodes is provided.

Note that here we have used the LHCII closed-system dynamics, valid
  only for short times, to partition it. As explained in
  Sec.~\ref{sec:quantumcom}, for the purpose of analysis one could
  alternatively use the less tractable open-system dynamics to obtain a
  partitioning that reflects the environment of the LHCII~\cite{MRLA08}.
  However, we argue that community partitioning, e.g.\ that based on the
  closed-system dynamics, is essential in devising approaches to simulating
  the full open-system dynamics.

\section{Discussion}
\label{sec:discussion}

We have developed methods to detect community structure in quantum systems,
thereby extending the purview of community detection from
classical networks to include quantum networks.
Our approach involves the development of a number of methods that
focus on different characteristics of the system and return
a community structure reflecting that specific characteristic. 
The variation of the quantum community structure with the property on which
this structure is based seems greater than for classical community
structures.

All our methods are based on the full unitary dynamics of the system,
as described by the Hamiltonian, and
account for quantum effects such as
coherent evolution and interference.
In fact phases are often fundamental to characterizing the system
evolution. For example,
Harel {\em et al.}~\cite{harel2012quantum} have shown that in light harvesting
complexes interference between pathways is important even at room temperature.
In our light harvesting complex example (see Sec.\ref{sec:lhcii}), 
the \emph{ab-initio} community structures provided by the
long-time measures propose consistent
communities that stretch across the lumenal and stromal layers of the complex, absent in
the structure proposed by the community.

Since we consider time evolution, the averaging time~$t$ acts
as a tuning parameter for the partitioning methods. In the case of transport
it transforms the method from a semi-classical approach ($t\to0$)
to a fully quantum-aware measure ($t\to\infty$),
For all times, the complexity of our algorithms scales polynomially in the
number of nodes $|\NNN|$, at worst $O(|\NNN|^3)$ if the diagonalization of
$H$ is required. This allows the study of networks with node numbers up the thousands and tens of thousands, which is appropriate for the real-world quantum networks currently being considered.

As with classical community structure, there are many possible
definitions of a quantum community. We restricted ourselves to two broad
classes based on transport and fidelity under coherent evolution, both
based on dynamics, though in the limits considered in this paper the
closenesses and thus quantum community structure can be expressed purely
in terms of static properties. We end by briefly discussing some other
possible definitions based on statics (the earliest classical community
definitions were based on statics~\cite{newman2009networks}).
The first type is based on some quantum state~$\ket{\psi}$, e.g. the ground state
of~$H$. We might wish to partition the network by repeatedly diving the
network in two based on minimally entangled bipartitions. This could be
viewed as identifying optimum communities for some cluster-based
mean-field-like simulation~\cite{luhmann2013bosonic} whose entanglement
structure is expected to be similar to~$\ket{\psi}$.
The second type is based directly on the spectrum of the Hamiltonian~$H$.
We might partition the Hilbert space into unions of the eigenspaces of~$H$
by treating the corresponding eigenvalues as 1D coordinates
and applying a traditional agglomerative or divisive clustering
algorithm on them.
Note that the resulting partitioning would normally not be in the position basis.

The use of community detection in quantum systems addresses an open
challenge in the drive to unite quantum physics and complex network
science, and we expect such partitioning, based on our definitions or
extensions such as above, to be used extensively in making the large
quantum systems currently being targeted by quantum physicists tractable to
numerical analysis. Conversely, quantum measures have also been shown to
add novel perspectives to classical network
analysis~\cite{sanchez2012quantum}.

\begin{acknowledgments}
We thank Michele Allegra, Leonardo Banchi, Giovanni Petri and Zoltan
Zimboras for fruitful discussions.
MF, THJ and JDB completed part of this study while visiting the Institute
for Quantum Computing, at the University of Waterloo.
PM acknowledges the Spanish MINCIN/MINECO project TOQATA (FIS2008-00784),
EU Integrated Projects AQUTE and SIQS, and HISTERA project DIQUIP\@.
THJ acknowledges the European Research Council under the European Union's
Seventh Framework Programme (FP7/2007-2013) / ERC Grant Agreement No.
319286, and the National Research Foundation
and the Ministry of Education of Singapore for support.
JDB acknowledges the Foundational Questions Institute (under grant
FQXi-RFP3-1322) for financial support.
All authors acknowledge the Q-ARACNE project funded by the Fondazione
Compagnia di San Paolo.
\end{acknowledgments}


\begin{thebibliography}{10}

\bibitem{girvan2002}
M.~Girvan and M.~E.~J. Newman.
\newblock Community structure in social and biological networks.
\newblock {\em Proceedings of the National Academy of Sciences},
  99(12):7821--7826, 2002.

\bibitem{porter2009communities}
Mason~A Porter, Jukka-Pekka Onnela, and Peter~J Mucha.
\newblock Communities in networks.
\newblock {\em Notices of the AMS}, 56(9):1082--1097, 2009.

\bibitem{fortunato2010community}
Santo Fortunato.
\newblock Community detection in graphs.
\newblock {\em Physics Reports}, 486(3):75--174, 2010.

\bibitem{rice1927identification}
Stuart~A Rice.
\newblock The identification of blocs in small political bodies.
\newblock {\em The American Political Science Review}, 21(3):619--627, 1927.

\bibitem{zachary1977information}
Wayne~W Zachary.
\newblock An information flow model for conflict and fission in small groups.
\newblock {\em Journal of anthropological research}, pages 452--473, 1977.

\bibitem{lusseau2004}
David Lusseau and Mark~EJ Newman.
\newblock Identifying the role that animals play in their social networks.
\newblock {\em Proceedings of the Royal Society of London. Series B: Biological
  Sciences}, 271(Suppl 6):S477--S481, 2004.

\bibitem{jonsson2006cluster}
Pall~F Jonsson, Tamara Cavanna, Daniel Zicha, and Paul~A Bates.
\newblock Cluster analysis of networks generated through homology: automatic
  identification of important protein communities involved in cancer
  metastasis.
\newblock {\em BMC bioinformatics}, 7(1):2, 2006.

\bibitem{pimm1979structure}
Stuart~L Pimm.
\newblock The structure of food webs.
\newblock {\em Theoretical Population Biology}, 16(2):144--158, 1979.

\bibitem{krause2003compartments}
Ann~E Krause, Kenneth~A Frank, Doran~M Mason, Robert~E Ulanowicz, and William~W
  Taylor.
\newblock Compartments revealed in food-web structure.
\newblock {\em Nature}, 426:282--285, 2003.

\bibitem{guimera2005functional}
Roger Guimera and Luis A~Nunes Amaral.
\newblock Functional cartography of complex metabolic networks.
\newblock {\em Nature}, 433(7028):895--900, 2005.

\bibitem{holme2003subnetwork}
Petter Holme, Mikael Huss, and Hawoong Jeong.
\newblock Subnetwork hierarchies of biochemical pathways.
\newblock {\em Bioinformatics}, 19(4):532--538, 2003.

\bibitem{flake2002self}
Gary~William Flake, Steve Lawrence, C~Lee Giles, and Frans~M Coetzee.
\newblock Self-organization and identification of web communities.
\newblock {\em Computer}, 35(3):66--70, 2002.

\bibitem{gauvain2013communities}
L.~{Gauvin}, A.~{Panisson}, and C.~{Cattuto}.
\newblock {Detecting the community structure and activity patterns of temporal
  networks: a non-negative tensor factorization approach}.
\newblock {\em arXiv:1308.0723}, August 2013.

\bibitem{allegra2012}
Michele Allegra and Paolo Giorda.
\newblock Topology and energy transport in networks of interacting
  photosynthetic complexes.
\newblock {\em Phys. Rev. E}, 85:051917, May 2012.

\bibitem{plenio2008dephasing}
Martin~B Plenio and Susana~F Huelga.
\newblock Dephasing-assisted transport: quantum networks and biomolecules.
\newblock {\em New Journal of Physics}, 10(11):113019, 2008.

\bibitem{renger2006}
J.~Adolphs and T.~Renger.
\newblock How proteins trigger excitation energy transfer in the {FMO} complex
  of green sulfur bacteria.
\newblock {\em Biophys. J.}, 91(8):2778--2797, 2006.

\bibitem{baezbook}
J.~C. {Baez} and J.~{Biamonte}.
\newblock {A Course on Quantum Techniques for Stochastic Mechanics}.
\newblock {\em arXiv:1209.3632}, September 2012.

\bibitem{faccin2013degree}
Mauro Faccin, Tomi Johnson, Jacob Biamonte, Sabre Kais, and Piotr Migda\l{}.
\newblock Degree distribution in quantum walks on complex networks.
\newblock {\em Phys. Rev. X}, 3:041007, Oct 2013.

\bibitem{zimboras2013quantum}
Zoltan Zimboras, Mauro Faccin, Zoltan Kadar, James Whitfield, Ben Lanyon, and
  Jacob Biamonte.
\newblock Quantum transport enhancement by time-reversal symmetry breaking.
\newblock {\em Sci. Rep.}, 3:2361, 2013.

\bibitem{johnson2014quantum}
Tomi~H Johnson, Stephen~R Clark, and Dieter Jaksch.
\newblock What is a quantum simulator?
\newblock {\em arXiv:1405.2831}, 2014.

\bibitem{ringsmuth2012}
AK~Ringsmuth, GJ~Milburn, and TM~Stace.
\newblock Multiscale photosynthetic and biomimetic excitation energy transfer.
\newblock {\em Nature Physics}, 8(7):562--567, 2012.

\bibitem{fleming10}
Mohan Sarovar, Akihito Ishizaki, Graham~R. Fleming, and K.~Birgitta Whaley.
\newblock {Quantum entanglement in photosynthetic light-harvesting complexes}.
\newblock {\em Nature Physics}, 6(6):462--467, April 2010.

\bibitem{IF12}
A.~Ishizaki and G.~R. Fleming.
\newblock Quantum coherence in photosynthetic light harvesting.
\newblock {\em Annu. Rev. Cond. Mat. Phys.}, 3(1):333--361, 2012.

\bibitem{CF09}
Yuan-Chung Cheng and Graham~R. Fleming.
\newblock Dynamics of light harvesting in photosynthesis.
\newblock {\em Annual Review of Physical Chemistry}, 60(1):241--262, 2009.
\newblock PMID: 18999996.

\bibitem{caruso09}
F.~Caruso, A.~W. Chin, A.~Datta, S.~F. Huelga, and M.~B. Plenio.
\newblock Highly efficient energy excitation transfer in light-harvesting
  complexes: The fundamental role of noise-assisted transport.
\newblock {\em The Journal of Chemical Physics}, 131(10):105106, 2009.

\bibitem{MRLA08}
M.~{Mohseni}, P.~{Rebentrost}, S.~{Lloyd}, and A.~{Aspuru-Guzik}.
\newblock {Environment-assisted quantum walks in photosynthetic energy
  transfer}.
\newblock {\em J. Chem. Phys.}, 129(17):174106, 2008.

\bibitem{banchi2013foster}
Leonardo Banchi, Gianluca Costagliola, Akihito Ishizaki, and Paolo Giorda.
\newblock An analytical continuation approach for evaluating emissionlineshapes
  of molecular aggregates and the adequacy of multichromophoricförster theory.
\newblock {\em The Journal of Chemical Physics}, 138(18):--, 2013.

\bibitem{rokhsar1991gutzwiller}
Daniel~S Rokhsar and BG~Kotliar.
\newblock Gutzwiller projection for bosons.
\newblock {\em Physical Review B}, 44(18):10328, 1991.

\bibitem{krauth1992gutzwiller}
Werner Krauth, Michel Caffarel, and Jean-Philippe Bouchaud.
\newblock Gutzwiller wave function for a model of strongly interacting bosons.
\newblock {\em Physical Review B}, 45(6):3137, 1992.

\bibitem{sheshadri1993superfluid}
K~Sheshadri, HR~Krishnamurthy, Rahul Pandit, and TV~Ramakrishnan.
\newblock Superfluid and insulating phases in an interacting-boson model:
  mean-field theory and the rpa.
\newblock {\em EPL (Europhysics Letters)}, 22(4):257, 1993.

\bibitem{luhmann2013bosonic}
Dirk-S{\"o}ren L{\"u}hmann.
\newblock Cluster gutzwiller method for bosonic lattice systems.
\newblock {\em Phys. Rev. A}, 87:043619, Apr 2013.

\bibitem{verstraete2008matrix}
Frank Verstraete, Valentin Murg, and J~Ignacio Cirac.
\newblock Matrix product states, projected entangled pair states, and
  variational renormalization group methods for quantum spin systems.
\newblock {\em Advances in Physics}, 57(2):143--224, 2008.

\bibitem{cirac2009renormalization}
J~Ignacio Cirac and Frank Verstraete.
\newblock Renormalization and tensor product states in spin chains and
  lattices.
\newblock {\em Journal of Physics A: Mathematical and Theoretical},
  42(50):504004, 2009.

\bibitem{wang2011numerically}
Haobin Wang, Ivan Pshenichnyuk, Rainer Härtle, and Michael Thoss.
\newblock Numerically exact, time-dependent treatment of vibrationally coupled
  electron transport in single-molecule junctions.
\newblock {\em The Journal of Chemical Physics}, 135(24), 2011.

\bibitem{cao2013multilayer}
Lushuai Cao, Sven Krönke, Oriol Vendrell, and Peter Schmelcher.
\newblock The multi-layer multi-configuration time-dependent hartree method for
  bosons: Theory, implementation, and applications.
\newblock {\em The Journal of Chemical Physics}, 139(13), 2013.

\bibitem{pan2013architercture}
Xiaowei Pan, Zhenfeng Liu, Mei Li, and Wenrui Chang.
\newblock Architecture and function of plant light-harvesting complexes {II}.
\newblock {\em Current Opinion in Structural Biology}, 23(4):515 -- 525, 2013.

\bibitem{novoderezhkin2005lhcii}
Vladimir~I. Novoderezhkin, Miguel~A. Palacios, Herbert van Amerongen, and Rienk
  van Grondelle.
\newblock Excitation dynamics in the {LHCII} complex of higher plants: Modeling
  based on the 2.72 {\aa} crystal structure.
\newblock {\em The Journal of Physical Chemistry B}, 109(20):10493--10504,
  2005.
\newblock PMID: 16852271.

\bibitem{fleming2009lhcii}
Gabriela~S. Schlau-Cohen, Tessa~R. Calhoun, Naomi~S. Ginsberg, Elizabeth~L.
  Read, Matteo Ballottari, Roberto Bassi, Rienk van Grondelle, and Graham~R.
  Fleming.
\newblock Pathways of energy flow in {LHCII} from two-dimensional electronic
  spectroscopy.
\newblock {\em The Journal of Physical Chemistry B}, 113(46):15352--15363,
  2009.
\newblock PMID: 19856954.

\bibitem{carlsson2010characterization}
Gunnar Carlsson and Facundo M{\'e}moli.
\newblock Characterization, stability and convergence of hierarchical
  clustering methods.
\newblock {\em The Journal of Machine Learning Research}, 99:1425--1470, 2010.

\bibitem{guimera2004modularity}
Roger Guimer\`a, Marta Sales-Pardo, and Lu\'is A.~Nunes Amaral.
\newblock Modularity from fluctuations in random graphs and complex networks.
\newblock {\em Phys. Rev. E}, 70:025101, Aug 2004.

\bibitem{good2010performance}
Benjamin~H Good, Yves-Alexandre de~Montjoye, and Aaron Clauset.
\newblock Performance of modularity maximization in practical contexts.
\newblock {\em Physical Review E}, 81(4):046106, 2010.

\bibitem{lancichinetti2011oslom}
Andrea Lancichinetti, Filippo Radicchi, Jos{\'e}~J Ramasco, and Santo
  Fortunato.
\newblock Finding statistically significant communities in networks.
\newblock {\em PloS ONE}, 6(4):e18961, 2011.

\bibitem{estrada2011community}
Ernesto Estrada.
\newblock Community detection based on network communicability.
\newblock {\em Chaos: An Interdisciplinary Journal of Nonlinear Science}, 21,
  2011.

\bibitem{estrada2009community}
Ernesto Estrada and Naomichi Hatano.
\newblock Communicability graph and community structures in complex networks.
\newblock {\em Applied Mathematics and Computation}, 214(2):500 -- 511, 2009.

\bibitem{meila2001random}
Marina Meila and Jianbo Shi.
\newblock A random walks view of spectral segmentation.
\newblock In {\em Eighth International Workshop on Artificial Intelligence and
  Statistics}, 2001.

\bibitem{Delvenne2010}
J.-C. Delvenne, S.~N. Yaliraki, and M.~Barahona.
\newblock Stability of graph communities across time scales.
\newblock {\em Proceedings of the National Academy of Sciences},
  107(29):12755--12760, 2010.

\bibitem{rosvall2011infomap}
Martin Rosvall and Carl~T Bergstrom.
\newblock Multilevel compression of random walks on networks reveals
  hierarchical organization in large integrated systems.
\newblock {\em PloS one}, 6(4):e18209, 2011.

\bibitem{pons2005computing}
Pascal Pons and Matthieu Latapy.
\newblock Computing communities in large networks using random walks.
\newblock In {\em Computer and Information Sciences-ISCIS 2005}, pages
  284--293. Springer, 2005.

\bibitem{hastie2001elements}
Trevor Hastie, Robert Tibshirani, and Jerome Friedman.
\newblock {\em The elements of statistical learning theory}.
\newblock Springer New York:, 2001.

\bibitem{newman2004pre}
M.~E.~J. Newman and M.~Girvan.
\newblock Finding and evaluating community structure in networks.
\newblock {\em Phys. Rev. E}, 69:026113, Feb 2004.

\bibitem{Newman2004}
M.~E.~J. Newman.
\newblock Fast algorithm for detecting community structure in networks.
\newblock {\em Phys. Rev. E}, 69:066133, Jun 2004.

\bibitem{newman2004finding}
Aaron Clauset, M.~E.~J. Newman, and Cristopher Moore.
\newblock Finding community structure in very large networks.
\newblock {\em Phys. Rev. E}, 70:066111, Dec 2004.

\bibitem{fortunato2007resolution}
Santo Fortunato and Marc Barthelemy.
\newblock Resolution limit in community detection.
\newblock {\em Proceedings of the National Academy of Sciences}, 104(1):36--41,
  2007.

\bibitem{nadakuditi2012graph}
Raj~Rao Nadakuditi and Mark~EJ Newman.
\newblock Graph spectra and the detectability of community structure in
  networks.
\newblock {\em Physical review letters}, 108(18):188701, 2012.

\bibitem{radicchi2013detectability}
Filippo Radicchi.
\newblock Detectability of communities in heterogeneous networks.
\newblock {\em Phys. Rev. E}, 88:010801, Jul 2013.

\bibitem{radicchi2014paradox}
Filippo Radicchi.
\newblock A paradox in community detection.
\newblock {\em EPL (Europhysics Letters)}, 106(3):38001, 2014.

\bibitem{ana2003normalized}
L.N.F. Ana and A.K. Jain.
\newblock Robust data clustering.
\newblock In {\em Computer Vision and Pattern Recognition, 2003. Proceedings.
  2003 IEEE Computer Society Conference on}, volume~2, pages II--128--II--133
  vol.2, 2003.

\bibitem{strehl2003cluster}
Alexander Strehl and Joydeep Ghosh.
\newblock Cluster ensembles---a knowledge reuse framework for combining
  multiple partitions.
\newblock {\em The Journal of Machine Learning Research}, 3:583--617, 2003.

\bibitem{danon2005}
Leon Danon, Albert D{\'i}az-Guilera, Jordi Duch, and Alex Arenas.
\newblock Comparing community structure identification.
\newblock {\em Journal of Statistical Mechanics: Theory and Experiment},
  2005(09):P09008, 2005.

\bibitem{lu2014chiral}
DaWei Lu, Jacob~D. Biamonte, Jun Li, Hang Li, Tomi~H. Johnson, Ville Bergholm,
  Mauro Faccin, Zolt{\'a}n Zimbor{\'a}s, Raymond Laflamme, Jonathan Baugh, and
  Seth Lloyd.
\newblock Chiral quantum walks.
\newblock {\em arXiv:1405.6209}, 2014.

\bibitem{eriksen2003modularity}
Kasper~Astrup Eriksen, Ingve Simonsen, Sergei Maslov, and Kim Sneppen.
\newblock Modularity and extreme edges of the internet.
\newblock {\em Physical review letters}, 90(14):148701, 2003.

\bibitem{weinan2008optimal}
E~Weinan, Tiejun Li, and Eric Vanden-Eijnden.
\newblock Optimal partition and effective dynamics of complex networks.
\newblock {\em Proceedings of the National Academy of Sciences},
  105(23):7907--7912, 2008.

\bibitem{godsil2011}
Chris Godsil.
\newblock Average mixing of continuous quantum walks.
\newblock arXiv:1103.2578 [math.CO], 2011.

\bibitem{traag2009}
V.~A. Traag and Jeroen Bruggeman.
\newblock Community detection in networks with positive and negative links.
\newblock {\em Phys. Rev. E}, 80:036115, 2009.

\bibitem{lancichinetti2008}
Andrea Lancichinetti, Santo Fortunato, and Filippo Radicchi.
\newblock Benchmark graphs for testing community detection algorithms.
\newblock {\em Phys. Rev. E}, 78:046110, Oct 2008.

\bibitem{harel2012quantum}
Elad Harel and Gregory~S. Engel.
\newblock Quantum coherence spectroscopy reveals complex dynamics in bacterial
  light-harvesting complex 2 (lh2).
\newblock {\em Proceedings of the National Academy of Sciences},
  109(3):706--711, 2012.

\bibitem{newman2009networks}
Mark Newman.
\newblock {\em Networks: an introduction}.
\newblock Oxford University Press, 2009.

\bibitem{sanchez2012quantum}
Eduardo S{\'a}nchez-Burillo, Jordi Duch, Jes{\'u}s G{\'o}mez-Garde{\~n}es, and
  David Zueco.
\newblock Quantum navigation and ranking in complex networks.
\newblock {\em Scientific reports}, 2, 2012.

\bibitem{jain1988algorithms}
Anil~K Jain and Richard~C Dubes.
\newblock {\em Algorithms for clustering data}.
\newblock Prentice-Hall, Inc., 1988.

\end{thebibliography}

\clearpage
\onecolumngrid
\appendix

\section{Definitions}
\subsection{Modularity}
Assume we have a directed, weighted graph (with possibly negative
weights) and self-links, described by a real adjacency matrix~$A$.
The element~$A_{ij}$ is the weight of the link from node~$i$ to
node~$j$.

The in- and outdegrees of node~$i$ are defined as
\begin{equation}
  k^{\text{in}}_i = \sum_j A_{ji}, \qquad
  k^{\text{out}}_i = \sum_j A_{ij}.
\end{equation}
For a symmetric graph~$A$ is symmetric and the indegree is equal to the
outdegree.
The total connection weight is
$m = \sum_i k^{\text{in}}_i = \sum_i k^{\text{out}}_i = \sum_{ij} A_{ij}$.

The community matrix~$C$ defines the membership of the nodes in
different communities. The element $C_{i \AAA}$ is equal to unity if $i \in
\AAA$, otherwise zero.%
\footnote{For a fuzzy definition of membership we could
require $C_{i \AAA} \geq 0$ and $\sum_\AAA C_{i \AAA} = 1$ instead.}
The size of a community is given by $|\AAA| = \sum_i C_{i \AAA}$.
For strict (non-fuzzy) communities we can define $C$ using
an assignment vector~$\sigma$ (the entries being the communities of
each node): $C_{i \AAA} = \delta_{\AAA, \sigma_i}$. This yields
$(C C^T)_{ij} = \delta_{\sigma_i, \sigma_j}$.

There are many different ways of partitioning a graph into communities.
A simple approach is to minimize the \emph{frustration} of the partition,
defined as the sum of the absolute weight of positive links between communities and negative links within
them:
\begin{equation}
F = -\sum_{ij} A_{ij} \delta_{\sigma_i, \sigma_j} = -\tr\left(C^T A C\right).
\end{equation}
Frustration is inadequate as a goodness measure for partitioning nonnegative graphs
(in which a single community containing all the nodes minimizes it).
For nonnegative graphs we can instead maximize another measure called \emph{modularity}:
\begin{equation}
Q = \frac{1}{m}\sum_{\AAA, ij} (A_{ij}-p_{ij}) C_{i \AAA} C_{j \AAA}
= \frac{1}{m}\tr\left(C^T (A-p) C\right),
\end{equation}
where $p_{ij}$ is the ``expected'' link weight from $i$ to~$j$, with~$\sum_{ij} p_{ij} = m$,
and is what separates modularity from plain frustration.
Different choices of the ``null model''~$p$ give different modularities.
Using degrees, we can define $p_{ij} = k^{\text{out}}_i k^{\text{in}}_j / m$.

For graphs with both positive and negative weights the usual definitions of degrees do not make much sense,
since usually negative and positive links should not simply cancel each other out.
Also, plain modularity will fail e.g.\ when~$m=0$.
This can be solved by treating positive and negative links separately~\cite{traag2009}.

\subsection{Hierarchical clustering}

All our community detection approaches share a common theme.
For each (proposed) community~$\AAA$
we have a goodness measure~$M_\AAA(t)$ that depends on the system Hamiltonian, the initial state, and~$t$.
This induces a corresponding measure for a partition~$\CC$:
\begin{align}
M_{\CC}(t) = \sum_{\AAA \in \CC} M_\AAA(t).
\end{align}
Using this, we define a function for comparing two partitions, $\CC$
and $\CC'$, which only differ in a single merge that combines
$\AAA$ and~$\BBB$:
\begin{align}
M_{\AAA, \BBB}(t)
= M_{\CC'}(t)-M_{\CC}(t)
= M_{\AAA \cup \BBB}(t) -M_\AAA(t)-M_\BBB(t).
\end{align}
We can make $M_{\AAA,\BBB}(t)$ into a symmetric closeness measure $c(\AAA,\BBB)$
by fixing the time~$t$ and normalizing it with~$|\AAA||\BBB|$.
Using this closeness measure together with the agglomerative
hierarchical clustering algorithm (as explained
in Sec.~\ref{sec:comdet}) we then obtain a community hierarchy.
The goodness of a specific partition in the hierarchy is given by its
modularity, obtained using the adjacency matrix given by
$A_{ij} = c(i,j)$.

The standard hierarchical clustering algorithm requires closeness to fulfill the \emph{monotonicity property}
\begin{align}
\label{eq:max}
\min(c(\AAA,\CCC), c(\BBB,\CCC)) \le c(\AAA \cup \BBB, \CCC) \le \max(c(\AAA,\CCC), c(\BBB,\CCC)).
\end{align}
for any communities~$\AAA, \BBB, \CCC$.
If this does not hold, we may encounter a situation where the
merging closeness sometimes increases, which in turn means that the
results cannot be presented as a dendrogram indexed by decreasing closeness.
The real downside of not having the monotonicity property, however, is stability-related.
The clustering algorithm should be stable, i.e. a small change in the system
should not dramatically change the resulting hierarchy.
Assume we encounter a situation where all the pairwise closenesses between a
subset of clusters $S = \{\AAA_i\}_i$
are within a given tolerance. A small perturbation can now change the
pair $\{\AAA,\BBB\}$ chosen for the merge.
If Eq.~\eqref{eq:max} is fulfilled, then 
the rest of $S$ is merged into the same new cluster during subsequent
rounds, and hence their relative merging order does not matter.

\subsection{Notation}

Let the Hamiltonian of the system have the spectral decomposition
$H = \sum_k \E_k \HPr_k$.
The unitary propagator of the system decomposes as
$U(t) = \mathrm{e}^{-\ii H t} = \sum_k e^{-i \E_k t} \HPr_k$.
We denote the state of the system at time~$t$ by
\begin{align}
\rho (t) =  U(t) \rho(0) U(t)^\dagger.
\end{align}
Sometimes we make use of the state obtained
by measuring in which community subspace $\VV_\AAA$ the quantum
state is located, and then discarding the result. The resulting state is
\begin{align}
\label{eq:measure}
\rho_\CC(t) &= \sum_{\AAA \in \CC} \SPr_\AAA \rho(t) \SPr_\AAA.
\end{align}
This state is normally not pure even if~$\rho(t)$ is.

The probability of transport from node~$b$ to node~$a$, the transfer matrix,
is given by the elements
\begin{align}
\TM_{ab}(t) = |\brackets{a}{U(t)}{b}|^2.
\end{align}
$\TM(t)$ is doubly stochastic, i.e.\ its rows and columns all sum up to unity.
We use $\sym{\TM} = (\TM+\TM^T)/2$ to denote its symmetrization.

The time average of a function $f(t)$ is denoted using~$\tave{f}(t)$:
\begin{align}
\tave{f}(t) = \frac{1}{t} \int_0^t f(t') \: \dd t'.
\end{align}
Now we have
\begin{align}
\tave{\TM}_{ab}(t)
&= \sum_{jk} \frac{1}{t} \int_0^t e^{-i(\E_j-\E_k)t'} \: \dd t' \brackets{a}{\HPr_j}{b} \brackets{b}{\HPr_k}{a}.
\end{align}
The $tH \ll 1$ and $t \to \infty$ limits of this average are
\begin{align}
\label{eq:T_ave_lims}
\notag
\tave{\TM}_{ab}(t \to 0) &= \delta_{ab}\left(1-\frac{t^2}{3}(H^2)_{aa}\right) +\frac{t^2}{3}|H_{ab}|^2 +O(t^3),\\
\tave{\TM}_{ab}(t \to \infty)
&= \sum_{jk} \delta_{jk} \brackets{a}{\HPr_j}{b} \brackets{b}{\HPr_k}{a}
= \sum_{k} |\brackets{a}{\HPr_k}{b}|^2.
\end{align}

The time average of the state of the system is given by
\begin{align}
\tave{\rho}(t)
= \sum_{jk} \frac{1}{t} \int_0^t e^{-i(\E_j-\E_k)t'} \: \dd t' \HPr_j \rho(0) \HPr_k.
\end{align}
It can be interpreted as the density matrix of a system that has
evolved for a random time, sampled from the uniform distribution on
the interval~$[0,t]$.
Again, in the short- and infinite-time limits this yields
\begin{align}
\label{eq:rho_ave_lims}
\notag
\tave{\rho}(t \to 0)
=& \rho(0) -\frac{it}{2} \left[H, \rho(0)\right]
+\frac{t^2}{3}\left(H \rho(0) H -\frac{1}{2}\left\{H^2, \rho(0)\right\} \right)
+O(t^3),\\
\tave{\rho}(t \to \infty)
=& \sum_{k} \HPr_k \rho(0) \HPr_k.
\end{align}

\section{Closeness measures}
\subsection{Inter-community transport}
\label{sec:S:mixing}

Considering the flow of probability during a continuous-time quantum
walk, let us investigate the \emph{change} in the probability of observing
the walker within a community:
\begin{align}
\T_\AAA (t)
= \frac{1}{2}\left| p_\AAA \left \{ \rho (t) \right \} - p_\AAA \left \{ \rho (0) \right \} \right|,
\end{align}
where
$p_\AAA \left \{ \rho \right \} = \tr \left( \SPr_\AAA \rho \right)$
is the probability of a walker in state~$\rho$ being found in
community~$\AAA$ upon a von Neumann-type measurement.\footnote{
Equivalently, $p_\AAA \left \{ \rho \right \}$ is
the norm of the projection (performed by projector $\SPr_\AAA$) of the
state $\rho$ onto the community subspace $\VV_\AAA$.}
A good partition should intuitively minimize this change, keeping the walkers as localized to the communities as possible.
$\T_\CC= \sum_{\AAA \in \CC}\T_{\AAA}$ is of course minimized by the trivial choice of a single
community, $\CC = \{\AAA\}$, and any merging of communities can only decrease~$\T_{\CC}$.
Therefore we have
$\T_{\AAA \cup \BBB}(t) \le \T_{\AAA}(t) +\T_{\BBB}(t)$.

The initial state~$\rho (0)$ can be chosen freely.
For a pure initial state $\rho(0) = \ket{\psi}\bra{\psi}$ we obtain
\begin{align}
\T_\AAA (t) = \frac{1}{2} \left| \bracket{\psi}{U^\dagger(t) \SPr_\AAA U(t)}{\psi} -\bracket{\psi}{\SPr_\AAA}{\psi}  \right|.
\end{align}
The change in inter-community transport is clearest when the process begins either entirely inside or entirely outside each community. Because of this, we choose the walker to be initially localized
at a single node $\rho (0) = \proj{b}$ and then, for symmetry, sum (or average)
$\T_\AAA (t)$ over all $b \in \NNN$:
\begin{align}
\label{eq:T}
\notag
\T_\AAA (t) &= \frac{1}{2} \sum_b  \left| \bracket{b}{U(t)^\dagger \SPr_\AAA U(t)}{b} -\bracket{b}{\SPr_\AAA}{b}  \right|\\
\notag
&= \frac{1}{2} \sum_b \left|\sum_{a \in \AAA} (\TM_{ab}(t) -\delta_{ab}) \right|\\
\notag
&= \frac{1}{2} \left(\sum_{b \in \AAA} \left|1 -\sum_{a \in \AAA} \TM_{ab}(t) \right| 
  +\sum_{b \notin \AAA} \left|\sum_{a \in \AAA} \TM_{ab}(t) \right| \right)\\
\notag
&= \frac{1}{2} \left(\sum_{a \notin \AAA,b \in \AAA} \TM_{ab}(t) 
  +\sum_{a \in \AAA, b \notin \AAA} \TM_{ab}(t)\right)\\
&= \sum_{a \in \AAA, b \notin \AAA} \frac{\TM_{ab}(t)+\TM_{ba}(t)}{2}
= \sum_{a \in \AAA, b \notin \AAA} \sym{\TM}_{ab}(t),
\end{align}
since $\TM(t)$ is doubly stochastic.
Now we have
\begin{align}
\T_{\AAA,\BBB}(t) = \T_{\AAA}(t) +\T_{\BBB}(t) -\T_{\AAA \cup \BBB}(t)
= 2 \sum_{a \in \AAA, b \in \BBB} \sym{\TM}_{ab}(t)
\end{align}
with $0 \le \T_{\AAA,\BBB}(t) \le 2 \min(|\AAA|, |\BBB|)$.
The short- and long-time limits of the time-averaged $\T_{\AAA,\BBB}(t)$
can be found using Eqs.~\eqref{eq:T_ave_lims}:
\begin{align}
\T_{\AAA, \BBB}^{t \to 0}
&= 2 \sum_{a \in \AAA, b \in \BBB}
\left( \delta_{ab} +\frac{t^2}{3}\left(|H_{ab}|^2 -\delta_{ab}(H^2)_{aa}\right) +O(t^3)\right),\\
\T_{\AAA, \BBB}^{t \to \infty}
&= 2 \sum_{a \in \AAA, b \in \BBB} \sum_k |(\HPr_k)_{ab}|^2.
\end{align}

\subsection{Intra-community fidelity}
\label{sec:S:coher}
Our next measure aims to maximize the ``similarity'' between the
evolved and initial states when projected to a community subspace.
We do this using the squared fidelity
\begin{align}
\F_\AAA (t) = F^2 \left \{ \SPr_\AAA \rho(t) \SPr_\AAA, \SPr_\AAA \rho(0) \SPr_\AAA \right \},
\end{align}
where $\SPr_\AAA \rho \SPr_\AAA$ is the projection of the state $\rho$ onto the subspace $\VV_\AAA$ and
\begin{align}
F \left \{ \rho , \sigma \right \} = \tr \left \{ \sqrt{ \sqrt{\rho} \sigma \sqrt{\rho} } \right \} \in  [0, \sqrt{\tr \{ \rho \} \tr \{ \sigma \}} ],
\end{align}
is the fidelity, which is symmetric between $\rho$ and $\sigma$.
If either $\rho$ or $\sigma$ is rank-1, their fidelity reduces to
$
F \left \{ \rho, \sigma \right \} = \sqrt{\tr \{\rho \sigma\}}
$.
Thus, if the initial state~$\rho(0)$ is pure, we have
\begin{align}
\F_\AAA (t) = \tr \left( \SPr_\AAA \rho(t) \SPr_\AAA \rho(0) \right).
\end{align}
This assumption makes
$\F_\CC(t)$ equivalent to the squared fidelity between $\rho_\CC(t)$ and a pure~$\rho(0)$:
\begin{align}
\label{eq:FX}
\notag
\F_\CC(t)
&= \sum_{\AAA \in \CC} \tr\left(\SPr_\AAA \rho(t) \SPr_\AAA \rho(0)\right)
= \tr\left(\rho_\CC(t) \rho(0)\right)\\
&= F^2\{\rho_\CC(t), \rho(0)\}
= F^2\{\rho(t), \rho_\CC(0)\},
\end{align}
and yields
\begin{align}
\label{eq:f2}
\notag
\F_{\AAA,\BBB}(t)
&= \F_{\AAA \cup \BBB}(t)-\F_\AAA(t)-\F_\BBB(t)\\
\notag
&= 2 \real \tr \left(\SPr_\AAA \rho(t) \SPr_\BBB \rho(0) \right)\\
&= 2 \sum_{a \in \AAA, b \in \BBB} \real \left(\rho_{ab}(t) \rho_{ba}(0) \right).
\end{align}
We use as the initial state the uniform superposition of all the basis states with arbitrary phases:
$\ket{\psi} = \frac{1}{\sqrt{n}}\sum_k e^{i \theta_k} \ket{k}$, which gives
\begin{align}
\F_{\AAA,\BBB}(t)
&= 
\frac{2}{n^2} \sum_{a \in \AAA, b \in \BBB} \sum_{xy}
\real\left(e^{i(\theta_x -\theta_y +\theta_b -\theta_a)} U_{ax} \overline{U_{by}}\right).
\end{align}
In this case the short-term limit does not yield anything interesting.
The long-time limit of the time-average of $\F_{\AAA,\BBB}(t)$ is
\begin{align}
\notag
\F_{\AAA, \BBB}^{t \to \infty}
&= \frac{2}{n^2} \sum_{a \in \AAA, b \in \BBB} \sum_{xy,k} \real \left(e^{i(\theta_x -\theta_y +\theta_b -\theta_a)}(\HPr_k)_{ax} (\HPr_k)_{yb} \right).
\end{align}
We may now (somewhat arbitrarily) choose all the phases~$\theta_k$ to be the same,
or average the closeness measure over all possible phases~$\theta_k \in [0, 2\pi]$.

\subsection{Purity}
The coherence between any communities $\CC = \{ \AAA,\BBB,\dots \}$ is completely destroyed
by measuring
in which community subspace $\VV_\AAA$ the quantum
state is located, see Eq.~\eqref{eq:measure}. If the measurement outcome is not revealed, the
purity of the measured state~$\rho_\CC(t)$ is,
due to the orthogonality of the projectors,
\begin{align}
\notag
\P_X(t) &= \tr\left(\rho_\CC^2(t)\right)
= \sum_{\AAA \in \CC} \tr\left((\SPr_\AAA \rho(t))^2\right)
= \sum_{\AAA \in \CC} \P_\AAA(t),
\end{align}
where
\begin{align}
\P_\AAA(t) &= \tr\left((\SPr_\AAA \rho(t) \SPr_\AAA)^2\right) = \tr\left((\SPr_\AAA \rho(t))^2\right).
\end{align}
If $\rho(t)$ is pure, we have (cf. Eq.~\eqref{eq:FX})
\begin{align}
\P_\CC(t)
= \sum_{\AAA \in \CC} \tr(\SPr_\AAA \rho(t) \SPr_\AAA \rho(t))
= F^2\{\rho_\CC(t), \rho(t)\}.
\end{align}

The change in purity of the state after a projective measurement
locating the walker into one of the communities is
\begin{align}
\notag
\P_{\AAA, \BBB}(t) &= \P_{\AAA \cup \BBB}(t) -\P_\AAA(t)-\P_\BBB(t)\\
\notag
&= 2\tr\left(\SPr_\AAA \rho(t) \SPr_\BBB \rho(t)\right)\\
&= 2 \sum_{a \in \AAA, b \in \BBB} | \rho_{ab} (t) |^2 \ge 0.
\end{align}

Again, we will use the initial state 
$\ket{\psi}~=~\frac{1}{\sqrt{n}}\sum_k e^{i \theta_k} \ket{k}$:
\begin{align}
\P_{\AAA,\BBB}(t)
&= 
\frac{2}{n^2} \sum_{a \in \AAA, b \in \BBB}
\left|\sum_{xy} e^{i(\theta_x-\theta_y)} U_{ax}(t) \overline{U_{by}(t)}\right|^2.
\end{align}
As with the fidelity-based measure, the short-time limit is uninteresting.
The long-time limit of the time-average of $\P_{\AAA,\BBB}(t)$ is
\begin{align}
\notag
\P_{\AAA,\BBB}^{t \to \infty}
&= 2 \sum_{a \in \AAA, b \in \BBB}
\left(|\bracket{a}{\tave{\rho}(\infty)}{b}|^2 +\sum_{k \neq m}
|\bracket{a}{\HPr_k \rho_0\HPr_m}{b}|^2\right)\\
&=
2 \sum_{a \in \AAA, b \in \BBB} \left( 
|\sum_{kxy} e^{i(\theta_x -\theta_y)} (\HPr_k)_{ax} (\HPr_k)_{yb}|^2 
+\sum_{k \neq m}|\sum_{xy} e^{i(\theta_x -\theta_y)} (\HPr_k)_{ax} (\HPr_m)_{yb}|^2
\right).
\end{align}

\begin{table*}
\renewcommand{\arraystretch}{1.5}
\begin{tabular}{lll|l}
method & initial state & limit &$A_{ab}$\\
\hline
\T & $\ket{j}$, summed over & before $t$-average &
$\frac{1}{2}\left(|U_{ab}(t)|^2 +|U_{ba}(t)|^2\right)$\\
& & $t \to 0$ &
$\delta_{ab} +\frac{t^2}{3} \left(|H_{ab}|^2 -\delta_{ab}(H^2)_{aa}\right) +O(t^3)$\\
& & $t \to \infty$ &
$\sum_{k} |(\HPr_k)_{ab}|^2$\\
\hline
\F & $\sum_j \ket{j}$ & before $t$-average &
$\sum_{xy} \real\left(e^{i(\theta_x -\theta_y +\theta_b -\theta_a)} U_{ax}(t) \overline{U_{by}(t)}\right)$\\
& & $t \to \infty$ &
$\sum_{x,y,k} \real \left((\HPr_k)_{ax} (\HPr_k)_{yb}\right)$\\
\hline
\F$^{\text{ph}}$ & $\sum_j e^{i \theta_j}\ket{j}$, & 
before $t$-average &
$\real \left(U_{aa}(t) \overline{U_{bb}(t)}\right) +\delta_{ab} \left(1-|U_{aa}(t)|^2\right)$\\
& phase-averaged & $t \to \infty$ &
$\sum_{k} (\HPr_k)_{aa} (\HPr_k)_{bb} +\delta_{ab} \left(1-\sum_{k} ((\HPr_k)_{aa})^2\right)$\\
\hline
\P & $\sum_j \ket{j}$ & before $t$-average &
$|\sum_{xy} e^{i(\theta_x-\theta_y)} U_{ax}(t) \overline{U_{by}(t)}|^2$\\
& & $t \to \infty$ &
$\left| \sum_{k,x,y}(\HPr_k)_{ax}(\HPr_k)_{yb}\right|^2 +\sum_{k \neq m} \left|\sum_{x,y}(\HPr_k)_{ax}(\HPr_m)_{yb}\right|^2$\\
\hline
\P$^{\text{ph}}$ & $\sum_j e^{i \theta_j}\ket{j}$, &
before $t$-average &
$1 +\delta_{ab} -\sum_{x} |U_{ax}(t)|^2 |U_{bx}(t)|^2$\\
& phase-averaged & $t \to \infty$ &
$1 +\delta_{ab} -\sum_x\left(\sum_{km} |(\HPr_k)_{ax}|^2 |(\HPr_m)_{bx}|^2 +\sum_{k
  \neq m}(\HPr_k)_{ax} (\HPr_k)_{xb} (\HPr_m)_{bx} (\HPr_m)_{xa} \right)$
\end{tabular}
\caption{Adjacency matrices, based on time-averaged measures. The closeness measure in each case is
$c(\AAA,\BBB) = \frac{2}{n^2 |\AAA||\BBB|} \sum_{a\in\AAA,b\in\BBB} A_{ab}$.
Note that
if $H$ has purely nondegenerate eigenvalues, then all the projectors are of
the form $\HPr_k = \outprod{\psi_k}{\psi_k}$, which makes some of the
$t\to\infty$ measures above identical.
For example $\T^{\infty}$ becomes the same as
$\F^{\infty, \text{ph}}$ outside the diagonal. This type of nondegeneracy
occurs e.g. when a small random perturbation is used to break the symmetries of~$H$.
}
\end{table*}

\end{document}